\documentstyle[12pt]{article}
\catcode`\@=11
\renewcommand{\thefootnote}{\fnsymbol{footnote}}

\begin{document}

\begin{titlepage}

{\hfill DFPD/94/TH/17}

\vspace{0.2cm}

{\hfill hep-th/9407091}

\vspace{0.2cm}

\vspace{1.2cm}

\centerline{{\bf NONPERTURBATIVE 2D GRAVITY, PUNCTURED SPHERES}}

\vspace{0.5cm}

\centerline{{\bf AND $\Theta$-VACUA IN STRING
THEORIES}\footnote[5]{P.A.M. and M.M. are Partly
supported by the European Community Research
Programme {\it Gauge Theories, applied supersymmetry and quantum
gravity}, contract SC1-CT92-0789\\
e-mail:  marchetti@padova.infn.it, matone@padova.infn.it}}

\vspace{1cm}

\centerline{{\bf G. Bonelli}\footnote{Supported by buona volont\`a},
{\bf P.A. Marchetti}
{\bf and M. Matone}}

\vspace{0.2cm}

\centerline{\it Department of Physics ``G. Galilei'' -
Istituto Nazionale di Fisica Nucleare}
\centerline{\it University of Padova}
\centerline{\it Via Marzolo, 8 - 35131 Padova, Italy}

\vspace{2.1cm}

\centerline{\bf Abstract}

\vspace{0.2cm}

We consider a model of 2D gravity with the coefficient of the
Euler characteristic having an imaginary part $\pi/2$.
This is equivalent to introduce a $\Theta$-vacuum structure in the genus
expansion whose effect is to convert the expansion into a series of
alternating signs, presumably Borel summable.   We show that the specific
heat of the model has a physical behaviour. It
can be represented nonperturbatively as a series
in terms of integrals over moduli spaces of punctured spheres and the
sum of the series can be rewritten as a unique integral over a suitable
moduli space of infinitely punctured spheres. This is an explicit
realization \`a la Friedan-Shenker of 2D quantum gravity.
We conjecture that the expansion in terms of punctures and the genus
expansion can be derived using the Duistermaat-Heckman theorem. We
briefly analyze expansions in terms of punctured spheres also for
multicritical models.

\end{titlepage}

\newpage

\setcounter{footnote}{0}

\renewcommand{\thefootnote}{\arabic{footnote}}

\noindent
{\bf 1.} The partition function of 2D quantum gravity is formally given by
\begin{equation}
Z_+=
\sum_{h=0}^\infty\int_{Met_h}{\cal D}ge^{-S(g)},
\qquad S(g)=\int_{\Sigma_h}\left(\lambda_1 \sqrt g +{\lambda_2\over 2\pi}
 \sqrt g R\right),
\label{oidhsd}\end{equation}
where $Met_h$ denotes the space of metrics on compact Riemann surfaces
$\Sigma_h$ of genus $h$ and $\lambda_1$, $\lambda_2$ are
bare parameters.
In this paper we discuss a generalization of (\ref{oidhsd}) based on the
analogy
with the theory of $\Theta$-vacua in
2D Yang-Mills theory.
In Abelian 2D gauge theories $\Theta$-vacua
can be introduced in the
path-integral formalism by adding the term
$i{\Theta\over 2\pi}\int F$ to the action
where $F$ is the curvature of the $U(1)$ gauge connection $A$.
The invariance of the gauge action without $\Theta$-term
under the transformation
$A\to -A$ guarantees the reality of the partition function for any real
value of $\Theta$.
In 2D gravity the
analogue of the $\Theta$-term is
\begin{equation}
{i\Theta\over 2\pi}\int_{\Sigma_h}
R\sqrt g=i\Theta \chi(\Sigma_h),
\label{0d2gd}\end{equation}
where $\chi(\Sigma_h)=2-2h$ denotes the Euler characteristic.
In 2D gravity, however, we have not an analogue of the symmetry
under $A\to -A$, but, since $\chi(\Sigma_h)$
is even, it follows that for
\begin{equation}
\Theta=k{\pi\over 2},\; k\in {\bf Z},
\label{oiodhqds}\end{equation}
the reality of the partition function is preserved.

Here we study a model of 2D quantum gravity with
$\Theta=\pi/2$.
The partition function of our model is given by
\begin{equation}
Z_-=
\sum_{h=0}^\infty\int_{Met_h}{\cal D}ge^{-S(g)+i{\Theta\over
2\pi}\int_{\Sigma_h}R\sqrt g}=
\sum_{h=0}^\infty(-1)^{1-h}\int_{Met_h}{\cal D}ge^{-S(g)},\quad
\Theta={\pi \over 2}.
\label{oidhsdaaa}\end{equation}
Expressions (\ref{oidhsd}) and (\ref{oidhsdaaa}) are purely
formal in two respect. First the integration measure on $Met_h$ has no
precise mathematical meaning: measures on infinite
dimensional spaces involve distributional completion of smooth
configuration spaces and it is not completely under control
how to define a distributional
completion of $Met_h$. Second the series
appearing in (\ref{oidhsd}) and (\ref{oidhsdaaa}) could diverge.
These are standard problems in QFT. The first one is faced with
renormalization theory.
According to \cite{DADFKKMBKKM} one introduces
a cut-off
$\epsilon$ by
replacing the integration over $Met_h$ by a sum of triangulations
of $\Sigma_h$ by equal size $k$-gons (of area $\epsilon$) and then
taking the  $\epsilon\to 0$ limit.
 A crucial point is that by Euler formula it follows that
under discretization the value of the
$\Theta$-term (\ref{0d2gd}) in the action
does not change and this guarantees the reality of the
discrete approximation of the partition function for $\Theta={k\pi/2}$,
$k\in{\bf Z}$.

Let $\lambda_1(\epsilon)$ and $\lambda_2(\epsilon)$ denote
the bare parameters in the regularized theory.
Under renormalization $\Theta$ is unchanged whereas the
renormalized
cosmological constant $\lambda_1^R$ and the
string coupling constant $g_s$ are given by
$$
\lambda_1^R=\lambda_1(\epsilon)-c/\epsilon,\quad
\lambda_2^R=\lambda_2(\epsilon)-{5\over 4} \log\epsilon,\quad
\log g_s\equiv\lambda_2^R.
$$
In terms of the renormalized parameters we have
\begin{equation}
Z_\pm=\sum_{h=0}^\infty
\left(g^{-1}_se^{i\Theta_\pm}\right)^{2-2h}\left(\lambda_1^R\right)^{{5\over 2}
(1-h)}Z_h,\qquad \Theta_+=0,\; \Theta_-=\pi/2.
\label{wouldberegularterms}\end{equation}
Let us define $t=\lambda_1^R \cdot g^{-4/5}_s$. It has
been conjectured  in \cite{mm} that there is
a nonperturbative definition of $Z_+$ through the matrix models such that
the ``specific heat''
$${\cal Z}_+(t)\equiv -Z_+''(t),$$
satisfies the Painlev\'e I (PI).
It has been proved
in \cite{FokasItsKitaev} that
by a suitable
modification the conjecture is true, although
the precise formulation of the matrix model is not completely
satisfactory (see
\cite{FDavid}\cite{SilvestrovYelkhovsky}\cite{FokasItsKitaev}\cite{rews}).

Let us briefly recall some basic facts about the solutions of
PI \cite{KapaevKitaev}
\begin{equation}
f^2(z)-{1\over 3}f''(z)=z.
\label{ddqds}\end{equation}
Its solutions can be characterized by five monodromy parameters
$s_l$, $l=1,\ldots,5$, satisfying
$$
s_{k+5}=s_k,\qquad s_{k+5}=i(1+s_{2+k}s_{3+k}),\qquad
k\in {\bf Z}.
$$
There are five one parameter families of solutions, $f_l(z)$, called
``simply truncated solutions'', characterized by the vanishing of
the monodromy parameter $s_{5-2l}$. In the sector
$$
\Omega_l\equiv \left\{-{2\pi\over 5}+{2\pi\over 5}l < \hbox{arg}\,
z< {2\pi\over 5}+{2\pi\over 5}l\right\},
$$
these solutions are characterized by the following asymptotic
\begin{equation}
f_l(z) =
\sqrt{|z|} e^{i\left({1\over 2} {\rm arg}\, z +\pi l\right)}
\sum_{n=0}^\infty b_n\left(
\sqrt{|z|} e^{i\left({1\over 2} {\rm arg}\, z +\pi l\right)}
\right)^{-5n}+f^{n.p.}_l(z),
\label{1}\end{equation}
where the coefficients $b_n$ satisfy the recursion relations
\begin{equation}
b_0=1\qquad
b_{n+1}={25n^2-1\over 24}b_n-{1\over 2}
\sum_{k=1}^n b_{n-k+1}b_k,
\label{oihxpedoihx}\end{equation}
and $f^{n.p.}_l(z)$ is a nonperturbative (``instanton'')
contribution. Along the rays $-{2\pi\over 5}+{2\pi\over 5}l$
and ${2\pi\over 5}+{2\pi\over 5}l$, the nonperturbative contributions
behave asymptotically as
\begin{equation}
f_l^{n.p.}\sim c_+|z|^{-{1\over 8}} s_{4-2l}
e^{ic |z|^{5\over4}+{\pi i\over 4}},
\quad
f_l^{n.p.}\sim -c_- |z|^{-{1\over 8}}
s_{1-2l}e^{ic |z|^{5\over4}-{\pi i\over 4}},
\label{oioxhowbis}\end{equation}
respectively, where $c_+$, $c_-$ and $c$ are nonvanishing
real constants given in \cite{KapaevKitaev}.
In any family of simply truncated solutions there is a special solution
characterized by $s_{5-2l}=0=s_{1-2l}$ called the ``triply truncated
solution''. It has poles only in the sector $\Omega_{l-1}$. Finally
there is a symmetry among the solutions of PI: if $f(z)$ is a solution
then also
\begin{equation}
\widehat f_k(z)\equiv
e^{-{4\pi i k\over 5}}f\left(e^{-{2\pi i k\over 5}}z\right),\qquad
k\in {\bf Z}/5{\bf Z},
\label{coiwjx}\end{equation}
is a solution. In terms of monodromy parameters the transformation
(\ref{coiwjx})
corresponds to $s_l\to s_{l-2k}$.

It has been shown that the nonperturbative definition
of the specific heat
${\cal Z_+}=-Z_+''$ obtained in the matrix model is in the family
of simply truncated solutions characterized by $s_5=0$, so that
(as one can check also by (\ref{1}))
$$
{\cal Z_+}(t)\to +\sqrt t, \qquad {\hbox{as}}\qquad
t\to +\infty,
$$
and the coefficients of its asymptotic expansion in $t$ are identified
with the $b_n$ in (\ref{oihxpedoihx}).
Substituting $g_s$ by $e^{\pm \pi i/2}g_s$ implies $t\to e^{\mp
2\pi i/5}t$ and perturbatively
\begin{equation}
Z_-(t)=Z_+\left(e^{\mp{2\pi i\over 5}}t\right),
\label{diuhd2iudh}\end{equation}
 so that
$\partial^2_tZ_-(t)=e^{\mp{4\pi i\over 5}}\partial^2_{t'}Z_+(t')$,
$t'=e^{\mp {2\pi i\over 5}}t$.
Given a solution ${\cal Z}_+$ of PI we set
\begin{equation}
{\cal Z}_-^{(1)}(t)\equiv e^{{4\pi i \over 5}}{\cal Z}_+\left(
e^{{2\pi i \over 5}}t\right),\qquad
{\cal Z}_-^{(2)}(t)\equiv e^{-{4\pi i \over 5}}{\cal Z}_+\left(
e^{-{2\pi i \over 5}}t\right).
\label{qidsguq}\end{equation}

By (\ref{coiwjx}) ${\cal Z}_-^{(1)}$ and
${\cal Z}_-^{(2)}$ are simply truncated solutions of PI characterized by
the vanishing of the monodromy parameters $s_2$ and $s_3$ respectively.
Their leading asymptotic behaviour is given by
\begin{equation}
{\cal Z}_-^{(k)}(t)\to -\sqrt t, \qquad {\hbox{as}} \;\,
t\to +\infty, \;\; k=1,2.
\label{questaequazionelalabelliamoanchesenonlarichiamiamo}\end{equation}
If one further imposes the physical reality condition for the
``specific heat'' ${\cal Z}$ of 2D gravity with $\Theta=\pi/2$, then
by (\ref{1}) and (\ref{oioxhowbis})
we find that ${\cal Z}_-^{(1)}={\cal Z}_-^{(2)}\equiv {\cal Z}_-$
is the triply truncated solution characterized by $s_2=s_3=0$. We expect
that this solution is the unique real one on the positive real axis
without ``instanton'' contributions $f_l^{n.p.}$.

\vspace{1cm}

\noindent
{\bf 2.} In this section we show that ${\cal Z}_-$
is strictly related to the theory of moduli space of punctured
Riemann spheres
compactified in the sense of Deligne-Knudsen-Mumford (DKM)
\cite{DeligneKnudsenMumford}.
In some sense, identifying $g_s$ as the coupling constant, we can think
about the genus expansion as a low-temperature expansion labelled by the
number of handles added to the sphere, whereas the expansion we will
describe can be thought of as a high-temperature expansion labelled by the
number of punctures inserted in the sphere.

It is not totally clear
to us how exactly one can derive
an expansion in terms of
 punctures from the genus expansion
(see later for comments on this point).
A general reason for this relation is nevertheless clear:
the (compactified) moduli space of
genus $h$ compact Riemann surfaces
$\overline{{\cal M}}_h\equiv \overline{{\cal M}}_{h,0}$
(we denote by ${\cal M}_{h,n}$ the moduli space of $n$-punctured Riemann
surfaces of genus $h$) has a boundary,
in the DKM compactification,
given as a sum of contributions of moduli spaces with one or two
punctures
\begin{equation}
\partial\overline{{\cal M}}_h=
c_{h,0}\overline{{\cal M}}_{h-1,2}+\sum_{k=1}^{[h/2]}
c_{h,k}\overline{{\cal M}}_{h-k,1}\times\overline{{\cal M}}_{k,1},
\label{uornavrlo}\end{equation}
in the sense of cycles on orbifolds. The coefficients $c_{h,j}$'s
are combinatorial factors.
Since each term in the RHS (\ref{uornavrlo}) has itself a boundary
given in terms of compactified moduli spaces of punctured surfaces
of lower genus and higher number of punctures, we can think to
iterate this procedure ending up with only compactified moduli spaces of
punctured spheres $\overline{{\cal M}}_{0,n}$.
In turn, $\overline{\cal M}_{0,n}$
has a boundary given in terms of compactified moduli spaces
of spheres with a lower number of punctures according to the recursion relation
\begin{equation}
\partial\overline{{\cal M}}_{0,n}=
\sum_{k=1}^{[n/2]-1}c_{0,k}^{(n)}\overline{{\cal M}}_{0,n-k}
\times\overline{{\cal M}}_{0,k+2}.
\label{ohdsq}\end{equation}
The coefficients $b_h$ of the asymptotic expansion of
${\cal Z}_\pm$ satisfy
the recursion relations (\ref{oihxpedoihx}) and
these relations appear to
be related to the decomposition (\ref{uornavrlo}) of the boundary
of $\overline{{\cal M}}_h$.
If one conjectures an analogous relation with (\ref{ohdsq}) for
 the coefficients ${\cal Z}_k$ of the expansion of
${\cal Z}(t)$ in terms of punctured spheres, then one is led to
the {\it Ansatz} considered in \cite{00}
\begin{equation}
{\cal Z}_n=d(n)\sum_{k=1}^{n-3}{\cal Z}_{k+2}{\cal Z}_{n-k}.
\label{iufcwocoihwe}\end{equation}
As shown in \cite{00} the solution of the PI
corresponding to (\ref{iufcwocoihwe}) is given in a neighborhood of
$t=0$ by the series
\begin{equation}
{\cal Z}(t)=\sum_{k=3}^{\infty}
{\cal Z}_k t^{5k-12},
\label{odqDQDQD}\end{equation}
with
\begin{equation}
d(n)={3\over (12-5n)(13-5n)},\qquad  {\cal Z}_3=-1/2.
\label{ueaw}\end{equation}
One can easily verify that such series
converges in a disk of finite radius
around the origin. Furthermore  the corresponding
solution of PI is characterized by the initial conditions
\begin{equation}
{\cal Z}(0)=0={\cal Z}'(0),
\label{gqwdkj}\end{equation}
and a computer simulation shows that
${\cal Z}(t)\sim -\sqrt{t}$
as $t\to +\infty$ (see figures) and it exhibits poles in the negative real
axis.
General arguments on the localization of poles
of  solutions
of the PI combined with bounds of the
coefficients of (\ref{odqDQDQD}) prove that the Borel sum of the series
(\ref{iufcwocoihwe})-(\ref{ueaw})
extends analytically the solution of PI to a neighborhood of the whole
positive real axis. Reality of the solution on the real axis
together with the asymptotic behaviour implies that the solution
(\ref{odqDQDQD}) can be indeed identified with the
triply truncated solution ${\cal Z}_-(t)$ ($s_2=s_3=0$) considered above.
Hence one can finally assert that the solution of the PI given on the
positive real axis by the Borel sum
\begin{equation}
{\cal Z}_-(t)={t^3}
\int_0^\infty dx e^{-x} \sum_{k=0}^\infty {\cal Z}_{k+3}{\left(t^5
x\right)^k\over k!},
\label{bergozza}\end{equation}
with the coefficients ${\cal Z}_k$'s given by (\ref{iufcwocoihwe}) and
(\ref{ueaw}) is exactly the ``specific heat'' of the 2D quantum gravity with
$\Theta=\pi/2$. In particular the $h^{th}$ coefficient of its asymptotic
expansion as $t\to +\infty$
is the partition function of such a  model of 2D gravity at genus
$h$.

If we apply to ${\cal Z}_-(t)$ the transformation (\ref{coiwjx})
corresponding to $t\to e^{{2\pi i\over 5}}t$, then we obtain a solution
${\cal Z}_+(t)$ of PI given by the triply truncated solution $s_5=s_4=0$
considered by \cite{mm}. As shown in
\cite{SilvestrovYelkhovsky}\cite{FokasItsKitaev} this solution
is not real on the real axis.
According to standard thermodynamics, if one defines (as done
here following \cite{mm}) the ``specific heat'' as the second
derivative of the free energy, it should be negative. One can check
that ${\cal Z}_-(t)$ is negative for all $t>0$ (see figures).
Notice instead that ${\cal Z}_+(t)$ is always positive
for sufficiently large $t$ which is an unphysical behaviour.

The asymptotic expansion
(as $t\to +\infty$) of ${\cal Z}_-$ is presumably Borel
summable. If this is the case then one can circumvent the problem of Borel
summability of the
asymptotic expansion in \cite{mm} by using the
relationship ${\cal Z}_+(t)=
e^{{4\pi i\over 5}}{\cal Z}_-\left(e^{{2\pi i\over 5}}t\right)$.

Let us comment on models with real valued $\Theta$ whose ``specific
heat'' is defined by
\begin{equation}
{\cal Z}_{\Theta}(t)\equiv e^{i{8\Theta\over5}}
{\cal Z}_+\left(te^{i{4\Theta\over5}}\right)=
-{d^2\over dt^2}Z_+\left(te^{i{4\Theta\over5}}\right).
\label{gdheiueoa}\end{equation}
The asymptotic expansion of the corresponding partition function is
given by
(\ref{oidhsdaaa}) for real $\Theta$.
By definition, ${\cal Z}_{\Theta}(t)$ satisfies the string equation
with complex parameters
\begin{equation}
f(t)^2-{1\over3}f''(t)=te^{i4\Theta},
\label{efwgfhfhfj}\end{equation}
which reduces to PI (\ref{ddqds}) only
for $\Theta={\pi\over2}k$, $k\in{\bf Z}$.

\vspace{1cm}

\noindent
{\bf 3.}
We now make explicit a closer connection between ${\cal Z}_-(t)$
and the structure of $\overline{{\cal M}}_{0,n}$ expressing the $n^{th}$
term in the series (\ref{odqDQDQD})
as an integral over $\overline{\cal M}_{0,n}$
and ${\cal Z}_-(t)$ as an integral on a suitable infinite
dimensional moduli space.

In a path integral approach we expect that
\begin{equation}
{\cal Z}_n=\int_{\overline{{\cal M}}_{0,n}}
{\omega^{(n)}}^{n-3}
e^{-{\cal S}_{(n)}}, \qquad \omega^{(n)}={\omega_{WP}^{(n)}\over \pi^2},
\label{runvabnarv}\end{equation}
where
${\cal S}_{(n)}$ is a suitable scalar action
and $\omega_{WP}^{(n)}$ is the Weil-Petersson two-form on
$\overline{{\cal M}}_{0,n}$.
 On the other hand it has been shown in \cite{00} that
one can write ${\cal Z}_n$ as the rational intersection number
\begin{equation}
{\cal Z}_n=\int_{\overline{{\cal M}}_{0,n}}
{\omega^{(n)}}^{n-4}\wedge\omega^{F_0}, \qquad n\geq 4,
\label{aarunvabnarv}\end{equation}
where $\omega^{F_0}$
is a closed two-form given in \cite{00} as a  linear combination
of the Poincar\'e dual of the divisors of
$\overline{{\cal M}}_{0,n}$ (we identify ${\cal Z}_k$ with
$Z^{F_0}_k$ in \cite{00}).

We now define the ``$q$-weighted'' moduli
space\footnote{``$q$-weighted'' moduli spaces have been
considered also in (\cite{GiventalKim}).}
of infinitely punctured
spheres $\overline{{\cal M}}_{0,\infty}(q)$ as follows:
let us consider the embedding
$i_n:\overline{{\cal M}}_{0,n}\to\overline{{\cal M}}_{0,n+1}$, $n>2$.
Then, for $q\in{\bf R_+}$, we define by inductive
limit
\begin{equation}
\overline{{\cal M}}_{0,\infty}(q)=
\coprod_{n=0}^{\infty}\left(
\overline{{\cal M}}_{0,n+3}\times [0,q^n]\right)/
\left(
\overline{{\cal M}}_{0,n},q^n\right)
\sim
\left(i_n
\overline{{\cal M}}_{0,n},0\right).
\label{gdjheuiekd}\end{equation}
Let $dy$ denote the Lebesgue measure on ${\bf R}$, assume the form
(\ref{aarunvabnarv}) for ${\cal Z}_n$ and define the indefinite rank
forms
\begin{equation}
\zeta_{\infty}=\left\{-{1\over2}+\sum_{k=1}^{\infty}
{{\omega^{(k+3)}}^{k-1}\wedge\omega^{F_0}
\over k!}\right\}\wedge dy,
\label{jkhncn}\end{equation}
\begin{equation}
{\cal S}_{\infty}=\left\{-{1\over2}+\sum_{k=1}^{\infty}
{{\omega^{(k+3)}}^{k-1}\wedge\omega^{F_0}
}\right\}\wedge dy.
\label{aaaaajkhncn}\end{equation}
Then by (\ref{odqDQDQD}), (\ref{bergozza}) and (\ref{gdjheuiekd})
it follows that
\begin{equation}
{\cal Z}_-(t)=t^3\int_0^\infty dx e^{-x}
\int_{\overline{{\cal M}}_{0,\infty}(t^5x)}
\zeta_{\infty},
\label{uyniym}\end{equation}
and, in a sufficiently small neighborhood of $t=0$, one can simply write
\begin{equation}
{\cal Z}_-(t)=t^3\int_{\overline{{\cal M}}_{0,\infty}(t^5)}
{\cal S}_{\infty}.
\label{uaaayniym}\end{equation}
Eqs.(\ref{uyniym})-(\ref{uaaayniym}) express nonperturbative
2D quantum gravity as
an integral on an infinite dimensional space which involves all
moduli spaces of punctured spheres.
The structure of this space and the crucial role of the
DKM compactification are at the basis of this explicit realization
of the Friedan-Shenker program \cite{FriedanShenker}.

\vspace{1cm}

\noindent
{\bf 4.}
Let us now consider the multicritical models, that is 2D quantum
gravity coupled to conformal matter \cite{mm}\cite{rews}.
In this section we will use the notations of \cite{00}.
The string equation for the $m^{th}$-model is
\begin{equation}
{m!\over (2m-1)!!}
\left\{
-1/2\partial^2_t+{\cal Z}^{F_{m-2}}(t)+
\partial_t^{-1}{\cal Z}^{F_{m-2}}(t)\partial_t
\right\}^m\cdot
1=t,
\label{strngqtn}\end{equation}
where ${\cal Z}^{F_{m-2}}$ denotes the ``specific heat''.
Consider the eq.(\ref{strngqtn}) with the initial conditions
$\left({\partial\over\partial t}\right)^k{\cal Z}^{F_{m-2}}(0)=0$,
for $k=0,\dots,2m-3$.
Since
the RHS of the string equation
vanishes for this choice of initial conditions it follows that
also $\left({\partial\over\partial t}\right)^{2m-2}
{\cal Z}^{F_{m-2}}(0)=0$.

These conditions imply that around $t=0$
\begin{equation}
{\cal Z}^{F_{m-2}}(t)=t^{-4(m+1)}\sum_{k=3}^\infty{Z}_k^{F_{m-2}}
t^{(2m+1)k},
\label{iujdqhfg}\end{equation}
where ${Z}_k^{F_{m-2}}$ are coefficients satisfying recursion relations
of the form
$$
{Z}_3^{F_{m-2}}={(-2)^{m-1}\over (2m-2)!!m!},
$$
\begin{equation}
{Z}_n^{F_{m-2}}=d(n,m)\left(\sum_{k=1}^{n-3}c(m;n,k)
{Z}^{F_{m-2}}_{n-k}
{Z}^{F_{m-2}}_{k+2}+\hbox{higher order terms}\right),\qquad n> m+1,
\label{udhqplk}\end{equation}
where $d(n,m)$ and $c(m;n,k)$ are positive factors.
The Taylor series (\ref{iujdqhfg}) is invariant
under the transformation $t\to te^{{2\pi i\over 2m+1}}$,
up to a $e^{{4\pi i\over 2m+1}}$ factor.
For $m$ even  (odd) it has alternating
(positive) signs, that is
\begin{equation}
{\rm sgn}\, {Z}_k^{F_{m-2}}=(-1)^{(m-1)k}.
\label{ihsgn}\end{equation}
This follows by the structure of the
string equation and eq.(\ref{udhqplk}).

It is instructive to analyze the $m=3$ model. The string
equation then reads
\begin{equation}
t={{\cal Z}^{F_1}}^3-{\cal Z}^{F_1}{{\cal Z}^{F_1}}''-
{1\over2}\left({{\cal Z}^{F_1}}'\right)^2+{1\over10}
{{\cal Z}^{F_1}}^{(4)}.
\label{itbckruc}\end{equation}
The expansion
${\cal Z}^{F_1}(t)=t^{-16}\sum_{k=3}^{\infty}{Z}^{F_1}_k
t^{7k}$ converges in a disk of finite radius
around $t=0$. Its coefficients are determined by the
recursion relations
$$
{Z}^{F_1}_3={1\over12},\quad
{Z}^{F_1}_4={5\cdot13\over2\cdot9\cdot 11\cdot (12)^3},
\quad
{Z}^{F_1}_n={10\over(7n-16)(7n-17)(7n-18)(7n-19)}\cdot
$$
\begin{equation}
\left(
{1\over2}\sum_{k=1}^{n-3}{Z}^{F_1}_{n-k}
{Z}^{F_1}_{k+2}(7k+8)(7n+7k-23)
 -\sum_{k=2}^{n-3}\sum_{j=1}^{k-1}{Z}^{F_1}_{n-k}
{Z}^{F_1}_{k-j+2}{Z}^{F_1}_{j+2}\right),
\quad n>4.
\label{jhnkljkm}\end{equation}

We can fully generalize to multimatrix models the results obtained in
\cite{00} for the $m=2$ case, i.e. one can prove that these are
``Liouville $F$-models''.
In fact, from eq.(\ref{udhqplk}), it
follows that the divisor structure
function is
\begin{equation}
F_{m-2}(n,k)=
{2(n-1)\over (n-4)!}{d(n,m)\left(c(m;n,k)Z^{F_{m-2}}_{n-k}
Z^{F_{m-2}}_{k+2}+\;{\rm higher}\;{\rm order}\;{\rm terms}\right)\over
a_{n-k}a_{k+2}},
\label{opejenfk}\end{equation}
for $n>1+m$,
while the lower ones are fixed by initial conditions (see \cite{00}).

Furthermore we argue that the following asymptotic expansion holds,
corresponding to our initial conditions:
\begin{equation}
{\cal Z}^{F_{m-2}}\sim
\sum_{k=0}^\infty
b^{(m)}_kt^{-{2m+1\over m}k+{1\over m}},\qquad {\rm as}
\;\, t\to e^{\pi i m}\cdot\infty
\label{aaaaicuhw}\end{equation}
where $b^{(m)}_0=(-1)^{m+1}$.

\vspace{1cm}

\noindent
{\bf 5.} We conclude the paper with some more speculative observations.

\vspace{0.5cm}

\noindent
{\it i}) The $\Theta$-vacuum ($\Theta=\pi/2$) can be
introduced in any string theory and it appears to convert a perturbative
genus expansion with definite signs for $\Theta=0$ to a genus expansion
with alternating signs. This might improve the Borel summability of the
perturbative series and therefore the $\Theta=\pi/2$ string theories
could be in general more tractable than the standard ones.

\vspace{0.5cm}

\noindent
{\it ii}) One can hope to implement more concretely the idea that the
recursion relations (\ref{oihxpedoihx}) and (\ref{iufcwocoihwe})
are related to the decomposition of $\partial\overline{{\cal M}}_h$
and $\partial\overline{{\cal M}}_{0,n}$, if the coefficients of the
genus expansion and the expansion in terms of punctured spheres, when
expressed as integrals on moduli spaces, can be evaluated by means of
a Duistermaat-Heckman (DH) theorem \cite{DuistermaatHeckman}.
Let $\omega$ be a symplectic form in a manifold $X$ of dimension $2n$
and $H$ an Hamiltonian on $X$; then, roughly speaking, DH tells that
integrals of the forms
$$
\int_X\omega^ne^{-H}
$$
only depend on the behaviour of the integrand near the critical points
of the flow of the Hamiltonian vector fields.
Equation (\ref{runvabnarv}) for $X=\overline{{\cal M}}_{0,n}$ or its
analogue in the genus expansion for $X=\overline{{\cal M}}_h$ are
exactly in the form that can be treated by DH, since the Weil-Petersson
2-form $\omega_{WP}$ is a symplectic form on $X$.
Hamiltonian vector fields are proportional to $\omega^{-1}_{WP}$
vanishing at $\partial X$ and this suggests that the critical points of
the Hamiltonian vector fields could be given by the boundary of the
relative moduli space. An application of DH then would imply that the
relevant integrals over moduli spaces reduce to integrals over their
boundaries.  Then the factorization property of the
Mumford forms \cite{BeilinsonManin}\cite{fay2} near
$\partial\overline{{\cal M}}_h$ would probably be
a key ingredient in the derivation of recursion relations
(\ref{oihxpedoihx}).

\vspace{0.5cm}

\noindent
{\it iii}) A feature of the genus expansion is that the first two terms
are distinguished with respect to the others. The reason is that the
Euler characteristic is positive for the Riemann sphere, zero for the
torus and negative for $h>1$. The path integral admits a semiclassical
limit ($g_s\to0$) only for $h>1$, since the Liouville equation appearing
in the classical limit does not admit solutions on positively curved
surfaces (see \cite{000} for a detailed discussion).
It is interesting to observe that instead only negative curved surfaces
appear in the expansion in terms of punctured spheres characterizing
the $\Theta=\pi/2$ model. Indeed this expansion precisely starts from
the sphere with three punctures, which is the ``minimum negatively curved
manifold''. This suggests that near the classical  limit factorizations
related to the DKM compactification together with unitarity requirement
could select among all complex values of the coefficient of the Euler
characteristic precisely the ones with imaginary part $\Theta=\pi/2$
(mod $\pi$).
In this respect it is intriguing to notice that phases naturally appear
in the factorization of the Mumford forms near
$\partial\overline{{\cal M}}_h$ (see {\it ii})).

Finally we note that our construction is strictly related to
topological field theories \cite{WittenDijkgraaf}\cite{BonoraXiong}
and the Kontsevich model
\cite{Kontsvevich}\cite{DiFrancescoItzyksonZuber}.

\vspace{1cm}

{\bf Acknoweldgements.} We would like to thank A. Bassetto, L. Bonora,
F. Fucito, C. Itzykson, K. Lechner, R. Nobili, M. Schlichenmaier, M. Tonin and
R. Zucchini for
suggestions and/or discussions. We thank Yu. Manin for bringing
ref.\cite{GiventalKim} to our attention.

\newpage

{\bf Figure Captions}

\vspace{0.5cm}

 Painlev\'e solution with boundary conditions $f(0)=0$, $f'(0)=0$.
$f$: solid line, $f'$: dotted line, $g(t)=-t^{1/2}$: dashed line.

\end{document}